\documentclass[11pt]{article}

\usepackage{geometry}
\usepackage[T1]{fontenc}
\usepackage{amssymb}
\usepackage{amsmath}
\usepackage{amsfonts}
\usepackage{authblk}
\usepackage{xcolor}
\usepackage[colorlinks,citecolor=blue,linktoc=all,linkcolor=blue,urlcolor=blue]{hyperref}
    
\author[1]{Bartosz Dziewit \thanks{\href{mailto:bartosz.dziewit@us.edu.pl}{bartosz.dziewit@us.edu.pl}}}
\author[1]{Joris Vergeest
\thanks{\href{mailto:jorisvergeest@hotmail.com}{jorisvergeest@hotmail.com}}}
\author[1]{Marek Zra\l ek
\thanks{\href{mailto:marek.zralek@us.edu.pl}{marek.zralek@us.edu.pl}}}
\affil[1]{Institute of Physics, University of Silesia,  Katowice, Poland}
\title{Flavor invariance of leptonic Yukawa terms in the 3HDM}
\begin{document}
\maketitle

\begin{abstract}
As an extension of the Standard Model (SM), the 3HDM (Three-Higgs-Doublet Model) defines additional relationships among the fermions. In the visible leptonic Yukawa sector of the 3HDM, we investigate the existence of flavor symmetries under discrete non-abelian groups up to order 1032. When the VEVs are not allowed to depart from their alignment imposed by the condition of minimal potential in the Higgs scalar sector, there will be severe mass degeneracies and a lack of lepton flavor mixing for any nontrivial flavor group. Yet we propose to study the Yukawa terms of the lepton sector for in case the VEV alignment would not be protected, hence leaving the VEV ratios as free parameters. Then, as it turns out, mass splitting for the charged leptons and the neutrinos is obtained although it is still impossible to obtain correct fits to the experimentally known lepton mass spectrum and to the PMNS matrix simultaneously.
\end{abstract}











\section{Introduction}
\label{sec:1}
For several decades, many extensions to the SM have been proposed, where major motivations include the accommodation of three fermion flavors with distinct masses and the incorporation of dark matter fields. A critical feature of models beyond the SM is their capacity to comprise three fermion families while correctly predicting their masses and mixing, or at least some of their relationships. To approach this long-standing flavor puzzle, it is crucial to describe how the global flavor symmetry (inherent in the SM) breaks down below the electroweak scale. A common strategy (that we also take) is to propose a particular finite flavor group $G$ and assign irreducible representations of $G$ (or one of its subgroups) to some or all field multiplets in flavor space. In doing so, it can be verified whether $G$ defines a discrete flavor symmetry of the Lagrangian. The Yukawa terms following from that symmetry define the fermionic mass matrices, from which the fermion masses can be derived as well as their mixing. These findings can be verified against experimental data. In the lepton sector (which is the scope of this study), there is accurate data on the charged lepton masses, partial knowledge about the neutrino masses, and (increasingly precise) information on lepton flavor mixing. If a symmetry group $G$ could be identified in accordance with lepton data, it might help to unravel the flavor puzzle.

The constraints implied by a $G$-symmetry will depend on the particle content added to the SM and on the chosen representation assignments. A common extension of the SM is formed by three right-handed neutrino fields, enabling the creation of massive neutrinos via the Dirac mechanism. It is not possible to consider the relationship between the masses of fermions with a single Higgs field, so the literature discusses proposals to expand the Higgs sector by including more fields, such as scalar fields (e.g., flavons) or Higgs scalar $SU(2)$ doublets.

The introduction of only one extra Higgs doublet to the SM forms a two-Higgs-doublet model (2HDM), which has been intensively studied (see e.g. \cite{BRANCO20121}). It was shown in \cite{Chaber:2018cbi} that the Yukawa sector of a 2HDM in a pure form (without additional flavons) is incompatible with the experimental lepton mass and mixing data whilst imposing a flavor symmetry of any $G$ with $|G| \le 1025$. Recently, in \cite{ChuliCentelles:2022ogma}, a 3HDM as an extension of the scotogenic model has successfully predicted lepton data and can accommodate dark matter fields under $\Sigma(81)$ flavor symmetry, where the neutrino masses are generated at one-loop. Another 3HDM, extended with a number of singlet scalars, correctly generates the SM fermion mass hierarchy through a sequential loop suppression mechanism \cite{Hern_ndez_2021}. Models with no extra Higgs doublets but extra flavons instead \cite{tribimaximalsmall} or under the restriction that the viable transformation subgroups are abelian \cite{Holthausen:2012wt} have been shown compatible with the PMNS and CKM data but are not conclusive on the lepton mass spectrum. In these studies, the candidate flavor group $G$ is explicitly broken down to two of its (abelian) subgroups, one for the charged lepton and one for the neutrino sector. For overviews of the vast number of approaches to lepton flavor symmetry research, we refer to \cite{Feruglio:2019ybq,King:2017guk,Ivanov_2017,Xing_2020,Chauhan:2023faf}.

Due to flavor symmetry group transformations of multi-Higgs models, only the Higgs potential and the Yukawa interaction part are non-invariant; the kinetic terms remain invariant. All allowed finite groups that leave the Higgs scalars potential invariant in the 3HDM have been completely classified \cite{Ivanov_2013, Ivanov:2014doa, Darvishi_2021}. The VEVs acquired after EWSB are determined by the global minimum of the scalar potential, where the vacuum alignment can be derived from the imposed flavor symmetry group. Such symmetry group should, after EWSB, leave the VEVs intact up to relabeling and a global phase. For that to happen, and depending on the VEVs (in particular their degeneracy) the symmetry group may need to be broken down to a flavor group preserving the Higgs vacua \cite{Gonz_lez_Felipe_2014}. This fact limits the feasibility of flavor groups of the full Lagrangian when the Higgs scalar doublets are assigned to a triplet. The no-go theorem proved in \cite{Gonz_lez_Felipe_2014} implies that any residual symmetry left in a flavor symmetry of the full Lagrangian of the quark sector that leaves the VEV alignment invariant up to relabeling and a global phase, will lead to degenerate quark masses and/or a block-diagonal CKM matrix. Similarly, in the lepton sector, masses and PMNS matrix consistent with experimental data cannot be obtained. The theorem holds for nHDMs with n>1 and for any occurring VEV alignments. It practically implies the absence of a phenomenologically viable remnant flavor symmetry in both the quark and lepton sectors, as exemplified for the groups $A_4$ and $S_4$ \cite{PhysRevD.87.055010}.

In the present study we show that in a 3HDM a nontrivial viable remnant symmetry $G$ is impossible not only due to constraints from the Higgs scalar sector described above, but it can also be derived from the Yukawa terms alone, for groups of order less than or equal to 1032. 
The VEVs are no longer constrained due to the scalar sector but instead transform as a flavor triplet. 
Only the Yukawa terms themselves remain invariant under $G$ and are assumed to generate the lepton mass matrices. This approach may be compared to the proposal of models involving dynamical EWSB, based on strong, dominant Yukawa interactions \cite{Bene__2009, Wetterich_2006, PhysRevD.21.3417}. Also in the analysis of the 2HDM in \cite{Chaber:2018cbi} it is assumed that the VEVs transform as 2D irreps under $G$. As common in multi-scalar models, such as in e.g. \cite{ChuliCentelles:2022ogma}, the inclusion of soft-breaking terms in the scalar potential may break the imposed flavor group explicitly at low energy, thus allowing the VEVs to deviate from their alignment and, in addition, such terms are needed to suppress FCNC. For the same reason the 3HDM proposed in \cite{Wyler:1979fe} is enriched with a fourth scalar field, not interacting with the fermions. Our aim is foremost to analyse the flavor symmetries in the lepton Yukawa terms, leaving the VEVs as free parameters, but insisting that the $SU(2)_L$ doublets remain intact under the flavor group. It then turns out that several groups give rise to lepton mass splitting. For example under $S_4$-symmetry, non-vanishing and non-degenerate charged lepton masses can be obtained, which would not be possible with the constraints on the VEVs, as shown in \cite{PhysRevD.87.055010}.

The Yukawa interaction terms considered in this stage of our study are limited to the Dirac mass terms $\mathcal{L}^l$ and $\mathcal{L}^{\nu}$ (defined below) for charged leptons and (left- and right-handed) neutrinos respectively, and the effective Weinberg operator $\mathcal{L}^M$ \cite{Weinberg:1979sa} for left-handed doublets.
We undertake a search of flavor symmetries for all groups $G$ with $|G| \le 1032$, where we focus strictly on the leptonic Yukawa terms of a 3HDM. The details of this 3HDM are defined in the next section. In section \ref{sec:3}, the procedure to identify symmetry groups is described, and in section \ref{sec:4}, the phenomenological viability of the symmetry groups is assessed.

\section{Flavor transformation of 3HDM leptonic Yukawa terms}
\label{sec:2}
 The Dirac mass terms and the Weinberg operator are defined as
\begin{eqnarray} \label{LAG1}
\mathcal{L}^l = && - (h^l_i)_{\alpha \beta} \overline{L}_{\alpha L} \tilde{\Phi}_i l_{\beta R} + \text{H.c.,}\label{Yukawa1} \\
\mathcal{L}^{\nu} =  && - (h^\nu_i)_{\alpha \beta} \overline{L}_{\alpha L} {\Phi}_i \nu_{\beta R}+ \text{H.c.,} \\
\mathcal{L}^M =  && -\frac{g}{M} (h^M_{i j})_{\alpha \beta}(\overline{L}_{\alpha L} \Phi_i)(\Phi^T_j (L_{\beta L})^c)+ \text{H.c.}
\end{eqnarray}
where summation over the lepton indices $\alpha, \beta = \text{e}, \mu, \tau$ and over the Higgs indices $i,j = 1,2,3$ is understood. $h^l_i$, $h^\nu_i$ and $h^M_{ij}$ are three-dimensional Yukawa matrices. $\Phi_i = (\phi^0_i \quad \phi^-_i)^T$ is an $SU(2)$ Higgs doublet; $\tilde{\Phi_i}=
(\phi^{- \star}_i -\phi^{0 \star}_i)^T$. The $L_{\alpha L}=(\nu_{\alpha L} \quad l_{\alpha L})^T$ are lepton doublets, and
$\bar{L}$ and $L^c$ denote adjoint and charge-conjugated lepton doublets, respectively.
Each mass matrix is linearly composed of three Yukawa matrices using the Higgs vacuum expectation values (VEVs) $v_i$:
\begin{eqnarray}
M^l && =-\frac{1}{\sqrt2} v^*_ih^l_i, \label{mass1}\\
M^{\nu} && =\frac{1}{\sqrt2}v_ih^{\nu}_i, \label{mass2}\\
M^M && =\frac{g}{M} v_iv_jh^M_{ij} \label{mass3}.
\end{eqnarray}

Since we assume that the left-handed leptons (of a given flavor) remain grouped in $SU(2)$ doublets, we study the invariance of two sums:
\begin{eqnarray} 
\text{either} \quad \mathcal{L}^l  +  \mathcal{L}^{\nu} \label{intterm1} \\
\text{or} \quad  \mathcal{L}^l  +   \mathcal{L}^M \label{intterm2}
\end{eqnarray}
when right-handed neutrinos are taken into account or not, respectively.
We will refer to these two expressions as the Dirac mass term and the Majorana mass term, respectively.

The triplets transform as defined by the three-dimensional representations of some candidate flavor group $G$. The Dirac (Majorana) mass term is $G$-symmetry if and only if Eq. \ref{intterm1} (Eq. \ref{intterm2}) is invariant under $G$.
For a candidate group $G$ we examine which of the expressions
\begin{eqnarray} \label{terms1}
&& \overline{L}_L A_L^\dagger\, (A_\Phi^* \tilde{\Phi})_i h^l_i\, A_{l R} \, l_R , \\
&& \overline{L}_L A_L^\dagger\, (A_\Phi \Phi)_i h^\nu_i\, A_{\nu R} \, \nu_R , \label{terms2}\\
&& \overline{L}_L A_L^\dagger\,(A_\Phi \Phi)_i (A_\Phi \Phi^T)_j) h^M_{ij} \, A^*_L \, L^c_R, \label{terms3}
\end{eqnarray}
(if any) remain invariant by the simultaneous matrix operators $A_L(g), A_{lR}(g), A_{\nu R}(g)$ and $A_\Phi(g)$ for all $g$ in $G$. The matrices $A(g)$ are defined by the group representations acting on the flavor triplets. Term \eqref{intterm1} can be invariant if $G$ has four (not necessarily different) representations $A_L$, $A_\Phi$, $A_{l R}$ and $A_{\nu R}$ such that both \eqref{terms1} and \eqref{terms2} remain invariant. Term \eqref{intterm2} can be invariant if $G$ has representations $A_L$, $A_{l R}$ and $A_\Phi$ such that both \eqref{terms1} and \eqref{terms3} remain invariant.

For details about the computational aspects of the invariance equations, see \cite{sym14091854}. To determine whether solutions $h^l_i$ and $h^\nu_i$ (or $h^l_i$ and $h^M_{ij}$) are in accordance with experimental data, we calculate the mass matrices using Eqs. \eqref{mass1},\eqref{mass2} and \eqref{mass3} using the two unitary matrices $U_l$ and $U_\nu$ that diagonalize the mass-squared matrices for charged leptons and for neutrinos, respectively:
\begin{eqnarray}\
U_l^\dagger(M^l M^{l \dagger}) U_l = (M_{diag}^l)^2, \\
U_\nu ^\dagger(M^\nu M^{\nu \dagger}) U_\nu = (M_{diag}^\nu)^2. \label{diag1}
\end{eqnarray}
The mass matrix $M^M$ is symmetric and diagonalized using unitary matrix $U_\nu$:
\begin{equation}\label{diag2}
U_\nu^T M^M U_\nu= M^M_d.
\end{equation}
The PMNS matrix is given by $U_l^\dagger U_\nu$.

\section{Finding flavor symmetry groups}
\label{sec:3}
The search for symmetries is based on a scan of finite groups $G$ of increasing order up to $|G|=1032$ using the SmallGroups library of GAP \cite{SmallGrp1.5.3,GAP4,sym14091854}. Matrix representations are obtained using the Repsn package of GAP \cite{Repsn}.
The conditions that we set for the representations in the model are:

\begin{itemize}
\item The transformation matrices in Eqs \eqref{terms1},\eqref{terms2},\eqref{terms3} are three-dimensional and unitary in order to conserve total lepton number and to ensure that  $\sqrt{\Sigma |v_i|^2}=(\sqrt{2}G_F)^{-1/2}$ GeV, where $G_F$ is the Fermi coupling constant of the SM.
\item The representations are irreducible. Reducible representations would lead to a combination of solutions also obtainable with (lower-dimensional) irreducible representations (see, e.g., \cite{ludl1}).
\item The solution of any invariance condition (Eqs. \eqref{terms1},\eqref{terms2} or \eqref{terms3}) must be unique; when multiple inequivalent solutions are found, the representation assignment is discarded, as to avoid the increase of the number of free parameters in the model.
\end{itemize}

We set no restrictions regarding the faithfulness of representations. It should be noted that an unfaithful representation of $G$ is isomorphic to a
factor group $F=G/H$, where $H$ is some normal subgroup of $G$. Both $H$ and $F$ are isomorphic to a subgroup of $U(3)$, but $F$ may or may not be isomorphic to a subgroup of $G$ (also, we do not require that $G$ is a subgroup of $U(3)$). For example, group $\Sigma (216\phi)$ has 6 faithful and 1 unfaithful 3D irreps. The character table of $G$ shows that for the unfaithful representation, 8 conjugacy classes have collapsed into the unit matrix. The union of those conjugacy classes (representing 54 group elements) defines $H$. The 648 elements of $G$ are thus partitioned into 12 parts. It was shown in Ref. \cite{ludl1} and in Refs.\cite{Jur_iukonis_2017, Hagedorn_2014} that these 12 parts define a group isomorphic to $A_4$, which is the factor group $F$, although $A_4$ is not a subgroup of $G$. As a consequence, a solution of the $G$-invariance condition may be based partly or entirely on factor groups of $G$, when some or all of the involved representations are unfaithful. The introduction of unfaithful representations can be seen as an implicit way to systematically account for transformations of different field triplets under different subgroups of $G$, as, e.g., explicitly performed in \cite{tribimaximalsmall}.

\section{Assessment of the compatibility with experimental lepton data}
\label{sec:4}

\subsection{Dirac mass term}
Out of the 1,152,579 groups that have 3 as a divisor of $|G|$, 9761 groups admit representation assignments that leave the Dirac mass term \eqref{intterm1} invariant. From these groups, 159,168 inequivalent solutions are obtained.
Each solution defines a pair of mass matrices $(M^l,M^\nu)$ (Eqs. \eqref{mass1}, \eqref{mass2}), which are functions of the VEVs. From these pairs follow the masses of the charged leptons and of the neutrinos, as well as the neutrino mixing angles (Eqs \eqref{diag1},\eqref{diag2}). Since the absolute strengths of the Higgs couplings are unknown in our model, only the mass proportions $m_\mu/m_e$ and $m_\tau /m_e$ can be calculated; the same holds for the mass proportions of the neutrino mass states. The VEVs ratios $v_2/v_1$ and $v_3/v_1$ are two complex parameters of the model.

To evaluate a solution, at first, it is checked if, for any choice of the VEVs, the calculated charged lepton mass ratios are consistent with experimental data. If so, then the focus is on $M^\nu$ to see if there exist VEVs to obtain viable charged lepton \textit{and} neutrino mass ratios. Finally, if any of the set of VEVs would give feasible mass ratios, also the computed neutrino mixing angles are evaluated.

The mass matrices $M^l$ (and similar for $M^\nu$) appear in three types (or textures). Mass matrices of type A are constructed with three Yukawa matrices, each having exactly one nonzero entry. For non-vanishing VEVs, the mass matrix appears to be monomial in all these cases. Mass matrices of type B occur when there are two nonzero entries in each of the three Yukawa matrices; they lead to a mass matrix with two nonzero entries in each row and two nonzero entries in each column. Type C mass matrices are obtained with Yukawa matrices having three or more nonzero entries; these form either a monomial mass matrix or a mass matrix with nonzero entries only.

\begin{table}
\caption{\label{tab:table1}Number of distinct flavor symmetric pairs $(M^l,M^\nu)$ and $(M^l,M^M)$, in table entries (X,Y) and (X,Y'), respectively. X,Y = A,B,C refers to the mass textures defined in the text.}
\centering
\begin{tabular}{lrrr|rrr}
\hline 
\hline
& A &  B & C &  A' &  B' & C'\\ \hline
A & 19333 & 0 & 4238 & 343 & 40 & 0  \\
B & 0 & 26 & 21 & 0 & 8 & 3 \\
C & 4198 & 21 & 131,331 & 72 & 0 & 0 \\ 
\end{tabular}
\end{table}

There are 23571 inequivalent solutions with $M^l$ of type A. Out of these, 19333 are accompanied by an $M^\nu$ of type A and 4238 of type C. Let us denote these types of solutions by (A,A) and (A,C), respectively (see Table \ref{tab:table1}). Solutions of type (A,B) do not occur. The smallest group generating a particular solution type is listed in Table \ref{tab:table2}.
The proportions $|v_2|/|v_1|$ and $|v_3|/|v_1|$ of all $M^l$ of type A are fixed by the experimentally known charged lepton mass ratios. If we label the Higgs fields such that the masses of the charged lepton mass states increase with increasing index, then $|v_2|/|v_1| = m_\mu/m_e \approx 206.7$ and $|v_3|/|v_1| = m_\tau /m_e \approx 3477$.
\begin{table}
\caption{\label{tab:table2}Smallest group $G$ giving rise to type (X,Y) or (X,Y'). X,Y = A,B,C refers to the mass textures defined in the text.}
\centering
\begin{tabular}{llc}
\hline 
\hline
Texture& $G$ &  GAP-ID  \\ \hline
(A,A) & $Z_7 \rtimes Z_3$ & [21,1]\\
(A,C) & $(Z_3 \times ((Z_3 \times Z_3) \rtimes Z_3)) \rtimes Z_3$ & [243,3] \\
(B,B) & $S_4$ & [24,12] \\
(B,C) & $Z_5 \times A_5$ & [300,22] \\
(C,A) & $(Z_3 \times ((Z_3 \times Z_3) \rtimes Z_3)) \rtimes Z_3$ & [243,3] \\
(C,B) & $Z_5 \times A_5$ & [300,22] \\
(C,C) & $Z_2 \times A_5$ & [120,35] \\
(A,A') & $(Z_4 \times Z_4) \rtimes Z_3$ & [48,3] \\
(A,B') & $((Z_4 \times Z_4) \rtimes Z_3) \rtimes Z_2$ & [96,64] \\ 
(B,B') & $((Z_3 \times Z_3 \times Z_3) \rtimes Z_3) \rtimes Z_2$ & [162,10] \\
(B,C') & $((Z_3 \times Z_3) \rtimes Z_3) \rtimes (Z_4 \rtimes Z_4)$ & [432,239]\\
(C,A') & $(Z_2 \times Z_2 \times Z_2 \times Z_2) \rtimes $ \\
 &   $\rtimes ((Z_3 \times Z_3 ) \rtimes Z_3)$ & [432,526]
\end{tabular}
\end{table}

A solution of type (A,A) implies that $m_2/m_1 \approx 206.7$ and $m_3/m_1 \approx 3477$, where $m_i$ denotes the mass of neutrino mass state $\nu_i$. We now check if these proportions are consistent with experimental observations.
In neutrino oscillation research the squared mass differences $\Delta m^2_{21}$ and $\Delta m^2_{32}$ have been measured
for normal ordering (NO) and inverted ordering (IO) of the neutrinos \cite{Workman:2022ynf}.
Let $\Delta m^2_{21} / \Delta m^2_{32}$ define $R_{NO}$ and $R_{IO}$ for NO and IO, respectively. Experimentally $R_{NO} = 0.0307 \pm 0.0009$ and $R_{IO} = -0.02969 \pm 0.0009$. $R_{NO}$ found by our model is 0.0036, and $R_{IO}$ is found -0.99. The discrepancy for NO is $\approx 30\sigma$ and for IO even much more.
It is thus concluded that solutions of type (A,A) can be compatible with the charged lepton masses but then are incompatible with the neutrino mass data.

In all (A,C) solutions, the $M^\nu$ matrix turns out to couple each left-handed flavor field to all three right-handed flavor fields via one Higgs doublet exclusively. The proportions of  $|v_i|$ (defined by $M^l$ of type A), will constrain $M^\nu$. Matrix $M^\nu M^{\nu\dagger}$ is diagonal with entries proportional to $|v_i|^2$, hence the derived neutrino mass proportions are fixed. We find $R_{NO} = 0.003532$ and $R_{IO} = -1.00$, respectively. So, it is concluded that the (A,C) solutions are incompatible with the neutrino mass data.

Inequivalent (C,A) solutions occur 4198 times in the group scan.
One variant of type C is a monomial mass matrix with the nonzero entries $d_i=|1+a_iy+b_iz|$, $i=(1,2,3)$. The complex coefficients $a_i$ and $b_i$ have modulus 1 and are defined by the Yukawa matrices. $y$ and $z$ are shorthand for $v_2/v_1$ and $v_3/v_1$. If there are (complex) parameters $y$ and $z$ such that $d_2/d_1\approx 207$ and $d_3/d_1 \approx 3477$, then the solution would be compatible with the experimental mass ratios of the charged leptons. For all $(y,z)$ giving the correct $d_2/d_1$, however, the ratio $d_3/d_1$ is close to $d_2/d_1$, about 17 times smaller than as known experimentally. Therefore, solutions (C,A), (C,B) and (C,C) are not in agreement with experimental data for this variant of type C. This can be proven analytically. There are other variants of type C that can only be analyzed numerically. Also, these variants do not admit a solution compatible with data from experiments.

Finally, there are 26 distinct (B,B) and 21 (B,C) solutions. 24 of these yield masses of the three charged leptons (and the neutrinos) proportional to (0,m,m), where $m \propto \sqrt{1+|y|^2+|z|^2}$. The remaining 23 solutions result in non-vanishing and non-degenerate masses, however with mass ratios much smaller than experimentally measured. Thus, all solutions of type (B,B) and (B,C) are in disagreement with experimental data.

\subsection{Majorana mass term}
The scan yields 1057 groups that admit representation assignments leaving the Majorana mass term \eqref{intterm2} invariant. The number of inequivalent solutions obtained from these groups is 466. The Majorana mass matrix $M^M$ is built of Yukawa matrices with either 1, 2 or 3 nonzero entries. The corresponding mass matrix types are denoted A', B', and C' in Table \ref{tab:table1}. The prime symbol distinguishes the notation of the Majorana mass term solutions from the one used for the Dirac mass term solutions.

Since $M^M$ is symmetric, a type A' matrix is diagonal. Its entries are proportional to the squares of the VEVs. The 383 invariance solutions involving such $M^M$ are either of type (A,A') or (C,A').
A type (A,A') solution is similar to the (A,A) solution of the Dirac mass term. The difference is that the squares of the neutrino mass ratios should be equal to the squares of the VEV proportions (these are derived from the experimentally measured charged lepton masses, as described above). The $R_{NO}$ and $R_{IO}$ values thus obtained are even farther off the experimental value compared to the value obtained for Dirac neutrinos.

The (C,A') solutions are discarded since an $M^l$ of type C gives wrong charged lepton masses, as shown above.

There are 40 type (A,B') solutions giving in all cases either $R_{NO}=139$ and $R_{IO}=-0.007$, or (partly or fully) degenerate neutrino masses.
Finally, 8 solutions of type (B,B') and 3 solutions of type (B,C') are encountered. These solutions can be discarded since the B-type $M^l$ returns no correct charged lepton mass ratios.

\section{Conclusions}
\label{sec:6}
We have studied the flavor invariance of the Dirac mass terms and the Majorana mass term (using the Weinberg effective operator) in the leptonic Yukawa sector of the 3HDM. It is known that after EWSB any nontrivial flavor group that stabilizes the VEV alignments in the Higgs scalar will imply degenerate lepton masses and/or lack of flavor mixing, contrary to experimental data. We find that when the constraints on the VEVs are removed, and the VEV ratios are taken as free parameters, mass splitting can be obtained in agreement with the experimentally known masses of either the charged leptons or neutrinos, but never simultaneously. In all cases the PMNS matrix is monomial. This has been verified for finite groups up to order 1032 after classifying all occurring mass matrices into three categories. Knowledge about (partly) viable flavor symmetries of the Yukawa sector, for VEVs deviating from their alignment might be relevant for the building of extended 3HDMs. In particular, several groups give rise to non-degenerate lepton masses, which would not be possible without lifting the constraints on the VEVs.

\section*{Acknowledgements}
This work has been supported in part by the Polish National Science Center (NCN) under grant 2020/37/B/ST2/02371 and the Research Excellence Initiative of the University of Silesia in Katowice.




\bibliographystyle{elsarticle-num}
\bibliography{bibliography.bib}






\end{document}